\documentclass[twocolumn, superscriptaddress]{revtex4}

\usepackage{amsmath}%,amssymb}

\usepackage{xcolor}
\usepackage{graphicx}

\usepackage[T1]{fontenc}
\usepackage[colorlinks,hyperindex]{hyperref}
\hypersetup
{
   colorlinks,%
   citecolor=blue,%
   linkcolor=blue,%
   urlcolor=blue,%
}
\usepackage[capitalise]{cleveref}

\mathchardef\mhyphen="2D

\newcommand\ph{\mathit{p\mhyphen h}}
\newcommand\logft{\mathrm{log}ft}

\makeindex

\begin{document}

\title{$\beta$-Delayed neutron spectroscopy of $^{133}$In}

\author{Z.~Y.~Xu}
\affiliation{Department of Physics and Astronomy, University of Tennessee,
Knoxville, Tennessee 37996, USA}
\author{M.~Madurga}
\affiliation{Department of Physics and Astronomy, University of Tennessee,
Knoxville, Tennessee 37996, USA}
\author{R.~Grzywacz}
\affiliation{Department of Physics and Astronomy, University of Tennessee,
Knoxville, Tennessee 37996, USA}
\affiliation{Physics Division, Oak Ridge National Laboratory, Oak Ridge,
Tennessee 37831, USA}
\author{T.~T.~King}
\affiliation{Department of Physics and Astronomy, University of Tennessee,
Knoxville, Tennessee 37996, USA}
\affiliation{Physics Division, Oak Ridge National Laboratory, Oak Ridge,
Tennessee 37831, USA}
\author{A.~Algora}
\affiliation{Instituto de F\'isica Corpuscular, CSIC-Universidad de Valencia,
E-46071, Valencia, Spain}
\affiliation{Institute for Nuclear Research (ATOMKI), H-4026 Debrecen, Bem ter
18/c, Hungary}
\author{A.~N.~Andreyev}
\affiliation{Department of Physics, University of York, North Yorkshire YO10
5DD, United Kingdom}
\affiliation{Advanced Science Research Center, Japan Atomic Energy Agency,
Tokai-mura, Japan}
\author{J.~Benito}
\affiliation{Grupo de F\'isica Nuclear and IPARCOS, Facultad de CC.\ F\'isicas,
Universidad Complutense de Madrid, E-28040 Madrid, Spain}
\author{T.~Berry}
\affiliation{Department of Physics, University of Surrey, Guildford GU2 7XH,
United Kingdom}
\author{M.~J.~G.~Borge}
\affiliation{Instituto de Estructura de la Materia, IEM-CSIC, Serrano 113 bis,
E-28006 Madrid, Spain}
\author{C.~Costache}
\affiliation{Horia Hulubei National Institute for Physics and Nuclear
Engineering, RO-077125 Bucharest, Romania}
\author{H.~De~Witte}
\affiliation{KU Leuven, Instituut voor Kern- en Stralingsfysica, B-3001 Leuven,
Belgium}
\author{A.~Fijalkowska}
\affiliation{Department of Physics and Astronomy, Rutgers University, New
Brunswick, New Jersey 08903, USA}
\affiliation{Faculty of Physics, University of Warsaw, PL 02-093 Warsaw, Poland}
\author{L.~M.~Fraile}
\affiliation{Grupo de F\'isica Nuclear and IPARCOS, Facultad de CC.\ F\'isicas,
Universidad Complutense de Madrid, E-28040 Madrid, Spain}
\author{H.~O.~U.~Fynbo}
\affiliation{Department of Physics and Astronomy, Aarhus University, DK-8000
Aarhus C, Denmark}
\author{A.~Gottardo}
\affiliation{IPN, IN2P3-CNRS, Universit\'e Paris-Sud, Universit\'e Paris Saclay,
91406 Orsay Cedex, France}
\author{C.~Halverson}
\affiliation{Department of Physics and Astronomy, University of Tennessee,
Knoxville, Tennessee 37996, USA}
\author{L.~J.~Harkness-Brennan}
\affiliation{Department of Physics, Oliver Lodge Laboratory, University of
Liverpool, Liverpool L69 7ZE, United Kingdom}
\author{J.~Heideman}
\affiliation{Department of Physics and Astronomy, University of Tennessee,
Knoxville, Tennessee 37996, USA}
\author{M.~Huyse}
\affiliation{KU Leuven, Instituut voor Kern- en Stralingsfysica, B-3001 Leuven,
Belgium}
\author{A.~Illana}
\affiliation{KU Leuven, Instituut voor Kern- en Stralingsfysica, B-3001 Leuven,
Belgium}
\affiliation{University of Jyv\"askyl\"a, Department of Physics, P.O. Box 35,
FI-40014, Jyv\"askyl\"a, Finland}
\author{\L.~Janiak}
\affiliation{Faculty of Physics, University of Warsaw, PL 02-093 Warsaw, Poland}
\affiliation{National Centre for Nuclear Research, 05-400 Otwock, \'swierk,
Poland}
\author{D.~S.~Judson}
\affiliation{Department of Physics, Oliver Lodge Laboratory, University of
Liverpool, Liverpool L69 7ZE, United Kingdom}
\author{A.~Korgul}
\affiliation{Faculty of Physics, University of Warsaw, PL 02-093 Warsaw, Poland}
\author{T.~Kurtukian-Nieto}
\affiliation{CENBG, Universit\'e de Bordeaux---UMR 5797 CNRS/IN2P3, Chemin du
Solarium, 33175 Gradignan, France}
\author{I.~Lazarus}
\affiliation{STFC Daresbury, Daresbury, Warrington WA4 4AD, United Kingdom}
\author{R.~Lic\u{a}}
\affiliation{ISOLDE, EP Department, CERN, CH-1211 Geneva, Switzerland}
\affiliation{Horia Hulubei National Institute for Physics and Nuclear
Engineering, RO-077125 Bucharest, Romania}
\author{R.~Lozeva}
\affiliation{Universit{\'e} Paris-Saclay, IJCLab, CNRS/IN2P3, F-91405 Orsay,
France}
\author{N.~Marginean}
\affiliation{Horia Hulubei National Institute for Physics and Nuclear
Engineering, RO-077125 Bucharest, Romania}
\author{R.~Marginean}
\affiliation{Horia Hulubei National Institute for Physics and Nuclear
Engineering, RO-077125 Bucharest, Romania}
\author{C.~Mazzocchi}
\affiliation{Faculty of Physics, University of Warsaw, PL 02-093 Warsaw, Poland}
\author{C.~Mihai}
\affiliation{Horia Hulubei National Institute for Physics and Nuclear
Engineering, RO-077125 Bucharest, Romania}
\author{R.~E.~Mihai}
\affiliation{Horia Hulubei National Institute for Physics and Nuclear
Engineering, RO-077125 Bucharest, Romania}
\author{A.~I.~Morales}
\affiliation{Instituto de F\'isica Corpuscular, CSIC-Universidad de Valencia,
E-46071, Valencia, Spain}
\author{R.~D.~Page}
\affiliation{Department of Physics, Oliver Lodge Laboratory, University of
Liverpool, Liverpool L69 7ZE, United Kingdom}
\author{J.~Pakarinen}
\affiliation{University of Jyv\"askyl\"a, Department of Physics, P.O. Box 35,
FI-40014, Jyv\"askyl\"a, Finland}
\affiliation{Helsinki Institute of Physics, University of Helsinki, P.O. Box 64,
FIN-00014, Helsinki, Finland}
\author{M.~Piersa-Si\l{}kowska}
\affiliation{Faculty of Physics, University of Warsaw, PL 02-093 Warsaw, Poland}
\author{Zs.~Podoly\'ak}
\affiliation{Department of Physics, University of Surrey, Guildford GU2 7XH,
United Kingdom}
\author{P.~Sarriguren}
\affiliation{Instituto de Estructura de la Materia, IEM-CSIC, Serrano 113 bis,
E-28006 Madrid, Spain}
\author{M.~Singh}
\affiliation{Department of Physics and Astronomy, University of Tennessee,
Knoxville, Tennessee 37996, USA}
\author{Ch.~Sotty}
\affiliation{Horia Hulubei National Institute for Physics and Nuclear
Engineering, RO-077125 Bucharest, Romania}
\author{M.~Stepaniuk}
\affiliation{Faculty of Physics, University of Warsaw, PL 02-093 Warsaw, Poland}
\author{O.~Tengblad}
\affiliation{Instituto de Estructura de la Materia, IEM-CSIC, Serrano 113 bis,
E-28006 Madrid, Spain}
\author{A.~Turturica}
\affiliation{Horia Hulubei National Institute for Physics and Nuclear
Engineering, RO-077125 Bucharest, Romania}
\author{P.~Van~Duppen}
\affiliation{KU Leuven, Instituut voor Kern- en Stralingsfysica, B-3001 Leuven,
Belgium}
\author{V.~Vedia}
\affiliation{Grupo de F\'isica Nuclear and IPARCOS, Facultad de CC.\ F\'isicas,
Universidad Complutense de Madrid, E-28040 Madrid, Spain}
\author{S.~Vi\~nals}
\affiliation{Instituto de Estructura de la Materia, IEM-CSIC, Serrano 113 bis,
E-28006 Madrid, Spain}
\author{N.~Warr}
\affiliation{Institut f\"ur Kernphysik, Universit\"at zu K\"oln, 50937 K\"oln,
Germany}
\author{R.~Yokoyama}
\affiliation{Department of Physics and Astronomy, University of Tennessee,
Knoxville, Tennessee 37996, USA}
\author{C.~X.~Yuan}
\affiliation{Sino-French Institute of Nuclear Engineering and Technology, Sun
Yat-Sen University, Zhuhai, 519082, Guangdong, China}

\date{today}

\begin{abstract}
   The decay properties of $^{133}$In were studied in detail at the ISOLDE Decay
   Station (IDS). The implementation of the Resonance Ionization Laser Ion
   Source (RILIS) allowed separate measurements of its $9/2^+$ ground state
   ($^{133g}$In) and $1/2^-$ isomer ($^{133m}$In). With the use of
   $\beta$-delayed neutron and $\gamma$ spectroscopy, the decay strengths above
   the neutron separation energy were quantified in this neutron-rich nucleus
   for the first time. The allowed Gamow-Teller transition
   $9/2^+\rightarrow7/2^+$ was located at 5.92 MeV in the $^{133g}$In decay with
   a $\logft=4.7(1)$. In addition, several neutron-unbound states were populated
   at lower excitation energies by the First-Forbidden decays of $^{133g,m}$In.
   We assigned spins and parities to those neutron-unbound states based on the
   $\beta$-decay selection rules, the $\logft$ values, and systematics.
\end{abstract}

\maketitle

\section{Introduction}

Doubly magic nuclei far from the stability line, such as $^{24}$O, $^{78}$Ni,
$^{100}$Sn, and $^{132}$Sn, have attracted tremendous interest in the last
decades. Their simple structures and imbalanced neutron-proton ratios provide a
testing ground to study the nuclear shell evolution as a function of isospin
both experimentally and theoretically, see, for example, Refs.\ \cite{24O_prl,
78Ni_nature, 100Sn_nature, 133Sn_kate, 131In_nature} and references therein. It
has been suggested that the residual $NN$ interactions are responsible for the
drift of single-particle orbitals and the modification of nuclear shell
structure at extreme neutron-proton ratios \cite{sorlin}. Since these phenomena
impact the decay properties of nuclei, nuclear $\beta$ decay is a viable probe
to study the shell evolution. Specifically, nuclear $\beta$ decay is extremely
sensitive to the occupation of proton and neutron orbitals. In the neutron-rich
nuclei, the predominant allowed Gamow-Teller (GT) transitions require $\Delta
L=0$ ($L$ refers to the orbital angular momentum of proton and neutron) between
neutron and proton spin-orbital partners in the same shell. The First-Forbidden
(FF) transitions, on the other hand, connect neutron and proton orbitals with
$\Delta L=1$ from neighboring shells. Due to their different selection rules,
the GT and FF operators transform the initial state of the parent nucleus into
diverse groups of final states in the daughter. Thus, measuring the decay
strengths as a function of excitation energy provides nuclear-structure
information in the parent and daughter nuclei. In addition, $\beta$ decays play
an important role in various nucleosynthesis processes. In the $r$-process, for
example, they compete with the rapid neutron-capture reaction and affect the
final elemental distribution \cite{rprocess-1, rprocess-2}. Owing to these
reasons, it is of particular interest to measure the $\beta$-decay properties in
the vicinity of $^{132}$Sn. Its proton and neutron shell closures at $Z=50$ and
$N=82$ define one of the strongest doubly magic cores on the neutron-rich side
of the nuclear chart \cite{132Sn_1, 132Sn_2, 133Sn_kate}, providing a reference
point to study nuclear structure with extreme neutron excess. Furthermore, the
proximity of the $r$-process path to $^{132}$Sn gives those decay properties key
impacts on the $r$-process abundance pattern near the mass number $A=130$ region
\cite{mumpower16}.

In this work, we studied the $\beta$ decay of $^{133}$In, a nucleus southeast of
$^{132}$Sn. Because of its substantial $Q_{\beta}$ window ($\sim13$ MeV), a
large number of states with different microscopic configurations can be
populated in the daughter $^{133}$Sn. The states below the neutron separation
energy were attributed to a single neutron outside the $^{132}$Sn core
\cite{hoff, 133Sn_kate, allmond}. The decay channels feeding those states can be
understood by transforming a neutron above $N=82$ into a proton below $Z=50$
(e.g., $\nu f_{7/2}\rightarrow\pi g_{9/2}$). The subsequent $\gamma$ decays had
been surveyed thoroughly by Piersa \textit{et al}.\ and Benito \textit{et al}.\
\cite{monika, benito}. The current study focused on the measurement of decay
strength above the neutron separation energy, which was less known
experimentally. The investigated states were highly excited because they were
dominated by the neutron or proton particle-hole (p-h) excitations with respect
to the $^{132}$Sn core. Promptly after the $\beta$ decays, neutrons were emitted
from those states, leaving the residual $^{132}$Sn in either the ground state or
excited states. Some of the experimental findings and their consequences were
highlighted in Ref.\ \cite{133In_prl}. This article explains in full detail the
experimental setup (\cref{exp}), the procedure to reconstruct the excitation
energies of the neutron-unbound states and their $\beta$-decay feeding
probabilities (\cref{analysis}), and the spin-parity assignments of these states
(\cref{spin}).

\section{Experiment}\label{exp}

The neutron-rich indium isotopes were produced at the Isotope Separator On-Line
(ISOLDE) facility at CERN \cite{isolde}. A 1.4-GeV proton beam was delivered by
the Proton Synchrotron Booster (PSB) and impinged on a tungsten solid neutron
converter \cite{converter} with an average current of 2 $\mu$A. Radioactive
isotopes, including $^{133}$In, were produced in a uranium carbide (UC$_x$)
target next to the neutron converter through neutron-induced fission. The indium
atoms were ionized using the Resonance Ionization Laser Ion Source (RILIS) at
ISOLDE \cite{rilis}. By using the narrow-band titanium-sapphire (Ti:Sa) laser at
RILIS \cite{narrowband}, selective ionization of $^{133}$In of either the ground
state or isomer can be achieved \cite{monika}. Following the General Purpose
Separator (GPS) \cite{isolde} separating the isotopes of interest based on the
mass-to-charge ratio, electrostatic quadrupoles transported the ion beam to IDS.
The beam was implanted into a movable tape at the center of a decay chamber.
After each proton pulse, the beam gate at ISOLDE was switched on for 300 ms for
continuous implantation. Then, the implantation was stopped for 300 ms before
the tape was rolled down to a shielded box to remove long-lived activities
originating from the daughter and granddaughter $\beta$ decays. The implantation
and tape move cycle provided a 600-ms time window to measure decay products from
the implanted $^{133}$In.

\begin{figure}[htbp]
   \centering
   \includegraphics[width=0.48\textwidth]{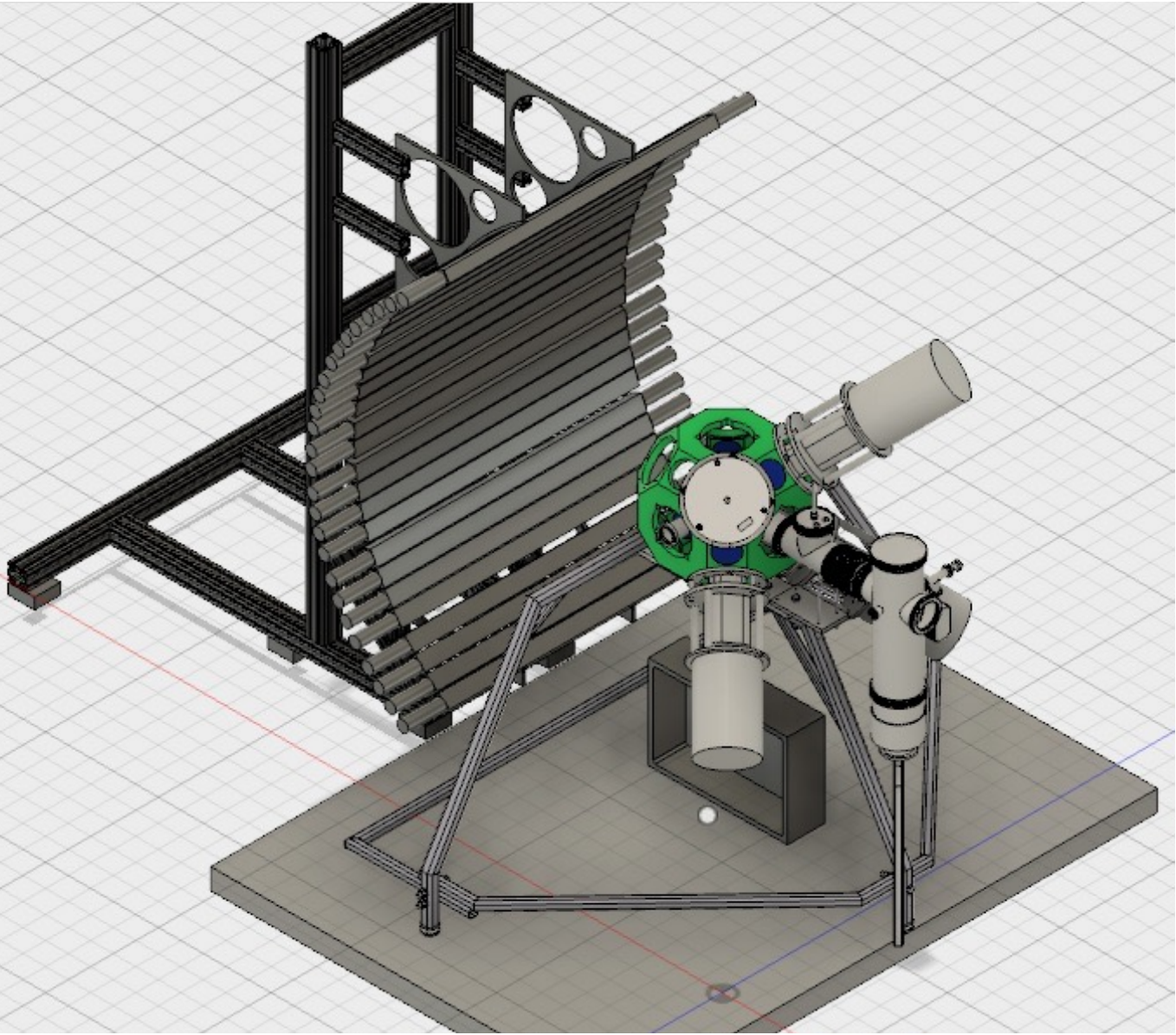}
   \caption{Schematic drawing of the experimental setup at IDS. The radioactive
   beam from ISOLDE was implanted on the tape at the center of the setup. The
   $\beta$-delayed $\gamma$ rays from $^{133}$In were detected by four HPGe
   clover detectors at backward angles relative to the beam direction. The INDiE
   array was placed on the other side to measure neutron spectroscopy following
   the $^{133}$In decay.
   }\label{ids}
\end{figure}

\Cref{ids} draws the detector configuration at IDS. Four high-purity germanium
(HPGe) clover detectors were placed closely outside the decay chamber to measure
$\beta$-delayed $\gamma$ radiation. The photopeak efficiency was 10\% and 3\% for
100-keV and 1-MeV $\gamma$ rays, respectively, including combining energy
deposition from Compton scattering inside all four crystals in each clover
(addback). The neutron energies $E_n$ were deduced from their time-of-flight
(TOF) measured by the IDS Neutron Detector (INDiE), an array similar to the
Versatile Array of Neutron Detectors at Low Energy (VANDLE) \cite{vandle,
miguelPRL2}. The detection setup consisted of 26 EJ-200 plastic scintillator
modules. Each module was $3\times6\times120$ cm$^3$ and had one photomultiplier
tube (PMT) coupled to each end. The modules were mounted in a custom-built
support frame describing an arch for a radius of 100 cm from the decay chamber.
The intrinsic neutron efficiency of each module was 35\% at 1 MeV \cite{vandle}.
The total solid angle covered by 26 modules was about 15\% of 4$\pi$. However,
only 22 out of 26 modules were used in the analysis due to the shadow from the
supporting frame of the decay chamber. The resulting solid angle was 12.6\% of
$4\pi$. Two plastic scintillators surrounding the implantation tape defined the
start signal of TOF. It provided average efficiency of up to $\sim$80\% for
$\beta$ particles. The neutron data were taken in the so-called triple
coincidence mode, requiring both PMTs of an INDiE module and one of the $\beta$
triggers to record an event. This way, the neutron detection threshold was
pushed down to 100 keV (or 5-keV$_{ee}$ energy loss in the detectors). The
traces of $\beta$ and INDiE signals were sampled by the 12-bit 250-MHz
digitizers. The sub-nanosecond time resolution was achieved using the algorithm
introduced in Ref.\ \cite{trace}. The FWHM of the $\gamma$-flash peak in the
obtained TOF spectrum was 1.5 ns. The actual neutron TOF distance between the
implantation point and INDiE modules was determined as 104.2(3) cm using the
online $\beta$ decay of $^{17}$N, which emits three fully resolved and
well-studied neutron lines \cite{17N}.

\section{Analysis of the neutron spectrum}\label{analysis}

\begin{figure*}[htbp]
   \centering
   \includegraphics[width=\textwidth]{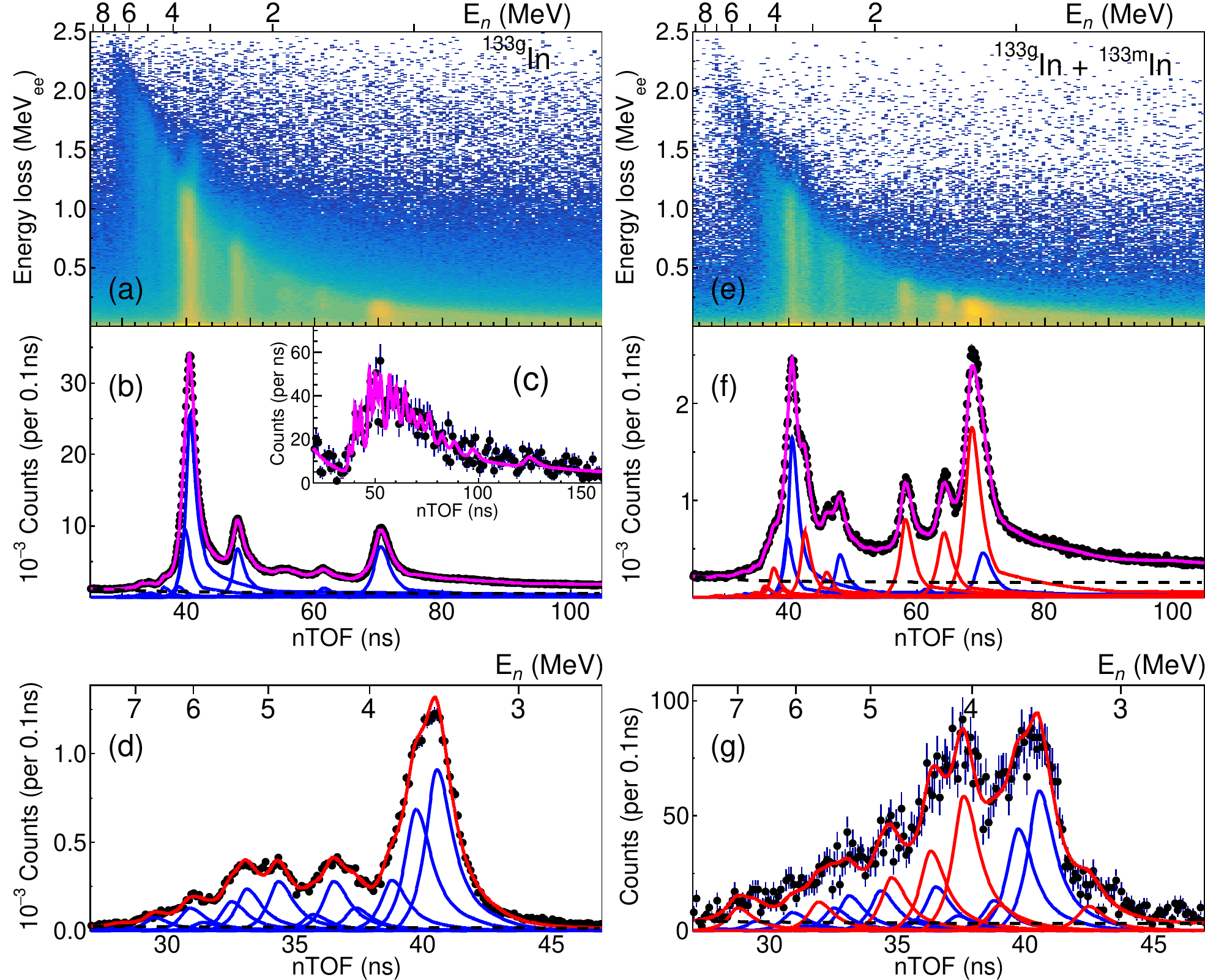}
   \caption{The neutron data taken in coincidence with the $^{133}$In $\beta$
   decay, with figures on the left corresponding to the pure ground-state decay
   and those on the right to an admixture of ground-state (40\%) and isomeric
   decays (60\%). Figures (a, e) plot the neutron TOF against their energy loss
   in INDiE, with the projections along TOF in Figs.\ (b, f), respectively.
   Figure (c) shows the ground-state neutron spectrum in coincidence with the
   4041-keV $\gamma$ decay in $^{132}$Sn. Figures (d, g) focus on the TOF of
   high-energy neutrons with $E_n>3$ MeV. They were made by projecting Figs.\
   (a, e), respectively, along TOF with energy loss greater than 1 MeV$_{ee}$.
   All the neutron TOF spectra were fitted by the neutron response functions
   (magenta), with the neutron peaks attributed to the ground-state and isomeric
   decays drawn in blue and red, respectively. The dashed line is the $\beta$
   and $\gamma$-ray background in the neutron TOF spectra.
   }\label{ntof}
\end{figure*}

Figures \ref{ntof} (a--g) present the neutron data taken in coincidence with the
$\beta$ decay of $^{133}$In. The figures on the left (a, b, c, and d) are taken
with the $^{133g}$In decay and those on the right (e, f, and g) with an
admixture of $^{133g}$In (40\%) and $^{133m}$In decays (60\%). When RILIS was
set on the $1/2^-$ isomer, there are still $^{133g}$In implanted into the decay
station at the same time, giving rise to the contamination peaks in the isomer's
neutron spectra.

Figures \ref{ntof} (a, e) show the collected neutron spectra in two-dimensional
(2D) histograms plotting the neutron TOF versus their energy loss in INDiE. The
neutron events are seen following the banana-shaped distribution in the
histograms. The neutron TOF spectra in Figs.\ \ref{ntof} (b, f) were made by
projecting the 2D histograms along the x-axis with energy loss greater than 5
keV$_{ee}$. Due to the simple structure of $^{133}$Sn and the $\beta$-decay
selection rules, only a few prominent neutron peaks were visible in the spectra.
Furthermore, no major peaks were observed in coincidence with $\gamma$ decays in
$^{132}$Sn, including the strongest $2^+_1\rightarrow0^+_{g.s.}$ transition, see
\cref{ntof} (c). This implied those $^{133}$Sn unbound states had a direct
feeding to the $^{132}$Sn ground state via neutron emissions. This lack of
neutron-$\gamma$ cascades was due to the first excited state ($2^+$) in
$^{132}$Sn being above 4 MeV, making it energetically impossible for most of the
neutron unbound states observed in the $^{133}$In decay. Nevertheless, there was
a small number of neutron emissions populating the $^{132}$Sn excited states.
Their numbers were estimated from the TOF spectra gated by the 4041-, 4352-, and
4416-keV $\gamma$ rays observed by the clover detectors. These $\gamma$ rays
correspond to the $2^+_1\rightarrow0^+_{g.s.}$, $3^-_1\rightarrow0^+_{g.s.}$,
and $4^+_1\rightarrow0^+_{g.s.}$ transitions in $^{132}$Sn, respectively.
According to Ref.\ \cite{benito}, they carry mostly the entire $\gamma$-decay
strength, via $\gamma$-$\gamma$ cascade, from an excited state to the ground
state, with a 5131-keV state being the sole exception. However, Ref.\
\cite{benito} reported an extremely weak ground-state feeding branching ratio
from this state. Thus, the error introduced by not considering this weak
neutron-$\gamma$ cascade was much smaller than the statistical uncertainties in
the analysis. The result showed about 7.0(5)\% of total neutron emissions going
to the $^{132}$Sn excited states. Their contribution was subtracted from the
total neutron activity to ensure the neutron intensities feeding the ground
state were extracted properly.

The neutron TOF spectra in Figs.\ \ref{ntof} (b, f) were fitted by a template
neutron response function. The procedure to determine the response function of
individual peaks is explained as follows. First, the TOF spectrum was simulated
for monoenergetic neutrons with GEANT4 \cite{geant4}, which took into account
all the neutron-scattering material at IDS and the time resolution of INDiE
modules. Second, the simulated profile was convolved with a Breit-Wigner style
distribution \cite{rmatrix} if a state had a sizeable width in energy (broad
resonance) greater than our resolution. The obtained response function was
verified using the $\beta$ decays of $^{49}$K and $^{17}$N, reproducing their
neutron spectra with only known peaks from the literature \cite{49K, 17N}.
Fitting the spectra of the $^{133}$In decays only involved "zero-width" peaks in
the response function. This indicated the observed resonances in $^{133}$Sn were
narrower than our energy resolution, which was about 80 and 250 keV for 1- and
3-MeV neutrons, respectively, at the minimum energy threshold (5 keV$_{ee}$).
Since the contribution from neutron-$\gamma$ cascades had been subtracted from
the fit, the peak intensity in the response function gave access to the neutron
intensities directly feeding the $^{132}$Sn ground state. The experimental
background, which was drawn as the dashed lines in \cref{ntof}, consisted of a
double-exponential decaying tail from fast $\beta$-decay electrons and a
constant plateau from random $\gamma$ rays.

\begin{table}
   \centering
   \caption{
      A list of neutron-unbound states identified in $^{133}$Sn. Those were
      derived from the neutron emissions directly feeding the $^{132}$Sn ground
      state. The excitation energies $E_{ex}$ were calculated from neutron TOF
      and the $S_\mathrm{n}=2.399(3)$ MeV \cite{ame2020}. The $I_{\beta}$ is the
      $\beta$-decay feeding probability given in \% per $\beta$ decay. A value
      with $\ast$ means the state decayed via competing neutron and $\gamma$
      channels, whereas that with $\dagger$ is likely a doublet at 6018 and 6088
      keV respectively. See text for details. The $\logft$ values were
      calculated using the half-lives from Ref.\ \cite{monika}, $\beta$-decay
      $Q_{\beta}$ from the atomic mass difference between $^{133}$In
      \cite{massIn133} and $^{133}$Sn \cite{ame2020}, and $E_{ex},\ I_{\beta}$
      from this work. Spins and parities $I^{\pi}$ were assigned tentatively
      based on the $\beta$-decay selection rules, $\logft$, and systematics.
   }\label{133Sn-table}
   \renewcommand{\arraystretch}{1.5}
   \setlength{\tabcolsep}{5pt}
   \begin{tabular*}{85mm}{@{\extracolsep{\fill}}clcllc}
      \hline
      \hline
      Parent      & $E_{ex}$ (keV)  & $I^{\pi}$                & $I_{\beta}$ (\%)   & $\logft$\\
      $^{133g}$In & 3562(18)        & $(11/2^-)$               & 12.0(8)$\ast$      & 5.8(1) \\
                  & 3923(27)        & $(7/2^-\sim11/2^-)$      & 1.8(1)$\ast$       & 6.5(1) \\
                  & 4123(35)        & $(7/2^-\sim11/2^-)$      & 0.6(1)$\ast$       & 6.9(1) \\
                  & 4236(36)        & $(7/2^-\sim11/2^-)$      & 0.8(1)             & 6.8(1) \\
                  & 4329(38)        & $(7/2^-\sim11/2^-)$      & 1.1(1)             & 6.6(1) \\
                  & 4906(55)        & $(7/2^-\sim11/2^-)$      & 9.5(3)             & 5.6(1) \\
                  & 5924(91)        & $(7/2^+)$                & 37.6(13)           & 4.7(1) \\
                  & 6068(96)        & $(7/2\sim11/2)$          & 14.0(7)$\dagger$   & 5.1(1) \\
                  & 6250(100)       & $(7/2\sim11/2)$          & 2.2(2)             & 5.9(1) \\
                  & 6550(120)       & $(7/2\sim11/2)$          & 0.3(1)             & 6.6(1) \\
                  & 6750(120)       & $(7/2\sim11/2)$          & 2.2(1)             & 5.7(1) \\
                  & 6950(130)       & $(7/2\sim11/2)$          & 0.3(1)             & 6.6(1) \\
                  & 7320(150)       & $(7/2\sim11/2)$          & 1.0(1)             & 5.9(2) \\
                  & 7700(160)       & $(7/2\sim11/2)$          & 1.1(1)             & 5.8(2) \\
                  & 7900(170)       & $(7/2\sim11/2)$          & 0.6(1)             & 5.9(2) \\
                  & 8300(190)       & $(7/2\sim11/2)$          & 0.2(1)             & 6.4(2) \\
                  & 8500(200)       & $(7/2\sim11/2)$          & 0.4(1)             & 5.9(2) \\
                  & 9100(230)       & $(7/2\sim11/2)$          & 0.3(1)             & 5.8(2) \\
                  %& 9800(270)       & $(7/2\sim11/2)$          & 0.3(1)             & 5.4(2) \\
      \hline
      $^{133m}$In & 3621(19)        & $(3/2^+)$                & 32.5(12)           & 5.5(1) \\
                  & 3794(23)        & $(1/2^+)$                & 11.9(5)            & 5.9(1) \\
                  & 4098(30)        & $(1/2^+,~3/2^+)$         & 13.6(5)            & 5.8(1) \\
                  & 4464(44)        & $(1/2^+,~3/2^+)$         & 0.9(2)             & 6.8(1) \\
                  & 4639(49)        & $(1/2^+,~3/2^+)$         & 0.9(2)             & 6.8(1) \\
                  & 5142(62)        & $(1/2^+,~3/2^+)$         & 4.3(2)             & 6.0(1) \\
                  & 5604(78)        & $(1/2^+,~3/2^+)$         & 11.7(5)            & 5.5(1) \\
                  & 6210(100)       & $(1/2,~3/2)$             & 1.4(3)             & 6.3(1) \\
                  & 6500(110)       & $(1/2,~3/2)$             & 5.7(4)             & 5.6(1) \\
                  & 6800(120)       & $(1/2,~3/2)$             & 2.3(2)             & 5.9(1) \\
                  & 7200(140)       & $(1/2,~3/2)$             & 1.1(2)             & 6.1(1) \\
                  & 8110(180)       & $(1/2,~3/2)$             & 0.6(2)             & 6.1(1) \\
                  & 9450(250)       & $(1/2,~3/2)$             & 0.5(2)             & 5.7(1) \\
      \hline
   \end{tabular*}
\end{table}

\begin{figure*}[htbp]
   \centering
   \includegraphics[width=.8\textwidth]{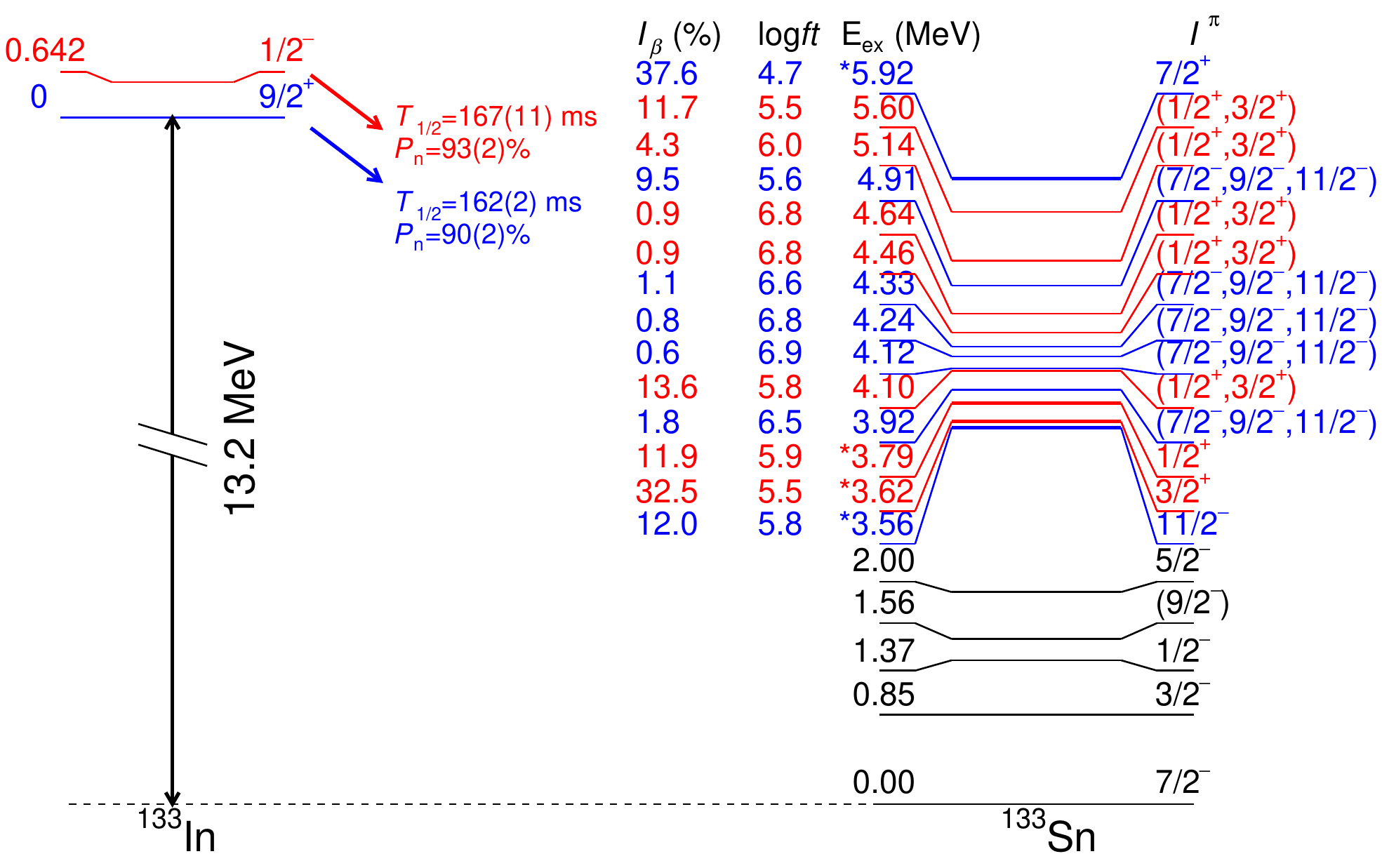}
   \caption{
      Constructed decay schemes of $^{133g}$In ($9/2^+$) and $^{133m}$In
      ($1/2^-$) between $S_\mathrm{n}=2.399$ MeV \cite{ame2020} and $E_{ex}=6.0$
      MeV, where isolated resonances were observed in $^{133}$Sn. Decays
      associated with the ground state and isomer are differentiated in blue and
      red, respectively. The states below the neutron separation energy are
      included for completeness \cite{monika, allmond}. Half-lives and
      neutron-emission probabilities ($P_n$) in $^{133}$In are taken from Refs.\
      \cite{monika, benito}. The excitation energy of the $1/2^-$ isomer is
      from Ref.\ \cite{massIn133}. The states with a solid spin-parity
      assignment are highlighted by ``$\ast$'' on their excitation energies.
   }\label{decayschemeP}
\end{figure*}

The results from a $\chi^2$-fitting analysis in \cref{ntof} (b, f) are
summarized in \cref{133Sn-table} and \cref{decayschemeP}. The numbers of peaks in
the response function to fit the ground-state and isomeric decays were 18 and 13,
respectively. All the ground-state peaks were included in the analysis of
isomeric decay due to contamination. Their contributions, as illustrated by the
blue peaks in \cref{ntof} (f) (and \cref{ntof} (g)), were determined by fixing
their relative intensities to the strongest peak at 41 ns with the ratios
obtained from the ground-state decay. The neutron peaks associated with the
isomeric decay were drawn in red in \cref{ntof} (f) (and \cref{ntof} (g)) to be
differentiated from the ground-state peaks. The excitation energies $E_{ex}$
were derived by summing the neutron kinetic energy, corrected by recoil energy,
with the neutron separation energy $S_\mathrm{n}=2.399(3)$ MeV in $^{133}$Sn
\cite{ame2020}. The experimental error combined the uncertainty in neutron TOF
centroid, flight distance, and neutron separation energy. In this analysis, the
number of detected $\beta$ decays $N_{\beta}$ was estimated from the number of
detected neutrons divided by the $\beta$-delayed neutron emission probability
($P_\mathrm{n}$), which is 90(2)\% and 93(2)\% for $^{133g}$In and $^{133m}$In
respectively \cite{benito}. Then, the decay probability $I_{\beta}$ of the state
was calculated by normalizing the neutron intensity to $N_{\beta}$. In cases
where there were $\gamma$ decays competing with neutron emissions, which will be
discussed in more detail later, $I_{\beta}$ included the contribution from
$\gamma$ decays. It is noted that all the $I_{\beta}$ in \cref{133Sn-table} were
calculated from the neutron emissions directly feeding the $^{132}$Sn ground
state. It is strictly correct for the states below 6.44 MeV in $^{133}$Sn. For
the states above 6.44 MeV, where the neutron-$\gamma$ cascade is energetically
possible, these $I_{\beta}$ should be regarded as the lower limits. The
ground-state (isomeric) decay $\logft$ values were extracted using the
$I_{\beta}$ in \cref{133Sn-table}, the $\beta$-decay half-life of $^{133g}$In
($^{133m}$In) from Ref.\ \cite{monika}, and the $Q_{\beta}=13.2$ MeV (13.8 MeV)
from the atomic mass difference between $^{133g}$In ($^{133m}$In)
\cite{massIn133} and $^{133}$Sn \cite{ame2020}.

A high-energy part of the neutron spectra in Figs.\ \ref{ntof} (b, f), i.e.,
$E_n>4$ MeV and TOF$<40$ ns, is more likely due to a continuum of strength
distribution rather than due to isolated resonances. There, a peak in the
response function should be regarded as a neutron quasi resonance
\cite{miguelPRL2}. The distribution of transitions in the continuum was inferred
by increasing the energy threshold of INDiE, resulting in better resolving power
in the TOF spectrum than the previously quoted values. For instance, Figs.\
\ref{ntof} (d, g) present the spectra with 1-MeV$_{ee}$ energy threshold for the
$^{133g,m}$In decays, respectively, showing better resolution for the
high-energy neutron peaks than \cref{ntof} (b, f). The centroids of the
high-energy neutron peaks were determined in \cref{ntof} (d, g) before they were
fixed in the fit of \cref{ntof} (b, f) to extract intensities together with
lower-energy neutron peaks.

\begin{figure}[htbp]
   \centering
   \includegraphics[width=0.5\textwidth]{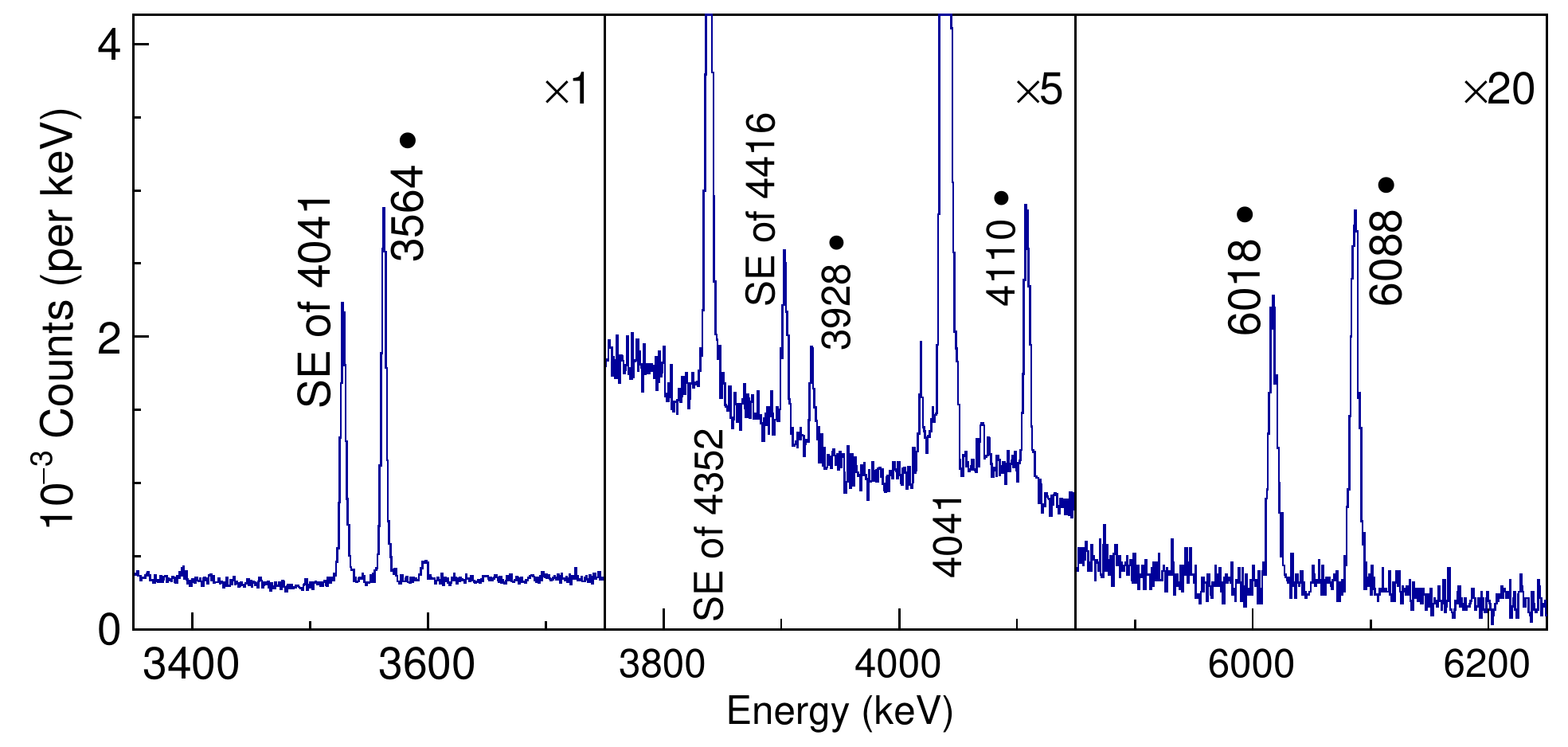}
   \caption{Portion of $\gamma$-ray spectra measured in coincidence with the
   ground-state decay of $^{133}$In. The candidate $\gamma$ deexcitations from
   the $^{133}$Sn neutron-unbound states are marked by ``$\bullet$'', to be
   distinguished from the $\gamma$ decays in $^{132}$Sn. SE stands for
   single-escape peak. See text for details.
   }\label{gamma}
\end{figure}

The observations of $\gamma$ decay from neutron-unbound states were reported
previously in $^{133}$Sn \cite{monika,vaquero17,benito}. Therefore, both
$\gamma$ and neutron intensities were measured in this work to ensure no
strength was missing in $I_{\beta}$. In the $^{133g}$In decay, five $\gamma$
peaks at 3564, 3928, 4110, 6018, and 6088 keV were found to have half-lives and
energies consistent with the corresponding neutron peaks in \cref{133Sn-table}.
These observed $\gamma$ rays agree with the previous $\beta$-decay study of
$^{133}$In using pure $\gamma$-ray spectroscopy \cite{monika, benito}.
\Cref{gamma} shows portions of $\gamma$-ray spectra in the relevant energy
ranges. Statistically, none of these $\gamma$ transitions were in coincidence
with any other $\gamma$ rays nor neutron emissions, suggesting they were
single-$\gamma$ transitions from a neutron unbound state to the ground state.
For the transitions at 3564, 3928, and 4110 keV, their $\gamma$ intensities were
added to the $I_{\beta}$ of 3561-, 3919-, and 4092-keV states in
\cref{133Sn-table}, respectively. Regarding the two $\gamma$ transitions around
6 MeV, their separation is only 70 keV, too close to be resolved by our neutron
detectors. Instead, a neutron peak was observed at $E_{n}=3642$ keV
($E_{ex}=6068$ keV) with TOF$\sim$39.5 ns, see \cref{ntof}(d). Thus, it was
presumed that the neutron peak consists of an unresolved doublet at 6018 and
6088 keV, respectively. The $I_{\beta}$ of the 6068-keV state corresponds to the
sum of the doublet. In contrast, no neutron-$\gamma$ competition was identified
following the $^{133m}$In decay. Below 6-MeV excitation energy, where $\beta$
feedings are strong, the $1/2^-$ isomer is expected to populate low-spin
positive-parity states in $^{133}$Sn via FF transitions. From those states, the
electromagnetic (EM) $E2$/$M1$ transitions to the $7/2^-$ ground state in
$^{133}$Sn are forbidden, and higher-order EM transitions ($M2$/$E3$) are too
slow to compete with neutron emissions. From a low-spin negative-parity state
fed by GT transitions at higher energy, the neutron-$\gamma$ competition is in
principle possible, similar to $^{133g}$In as discussed above. However, no such
candidates were found due to the combined effect of smaller $\beta$ feeding and
a limited number of implanted samples.

\begin{figure}[htbp]
   \centering
   \includegraphics[width=0.5\textwidth]{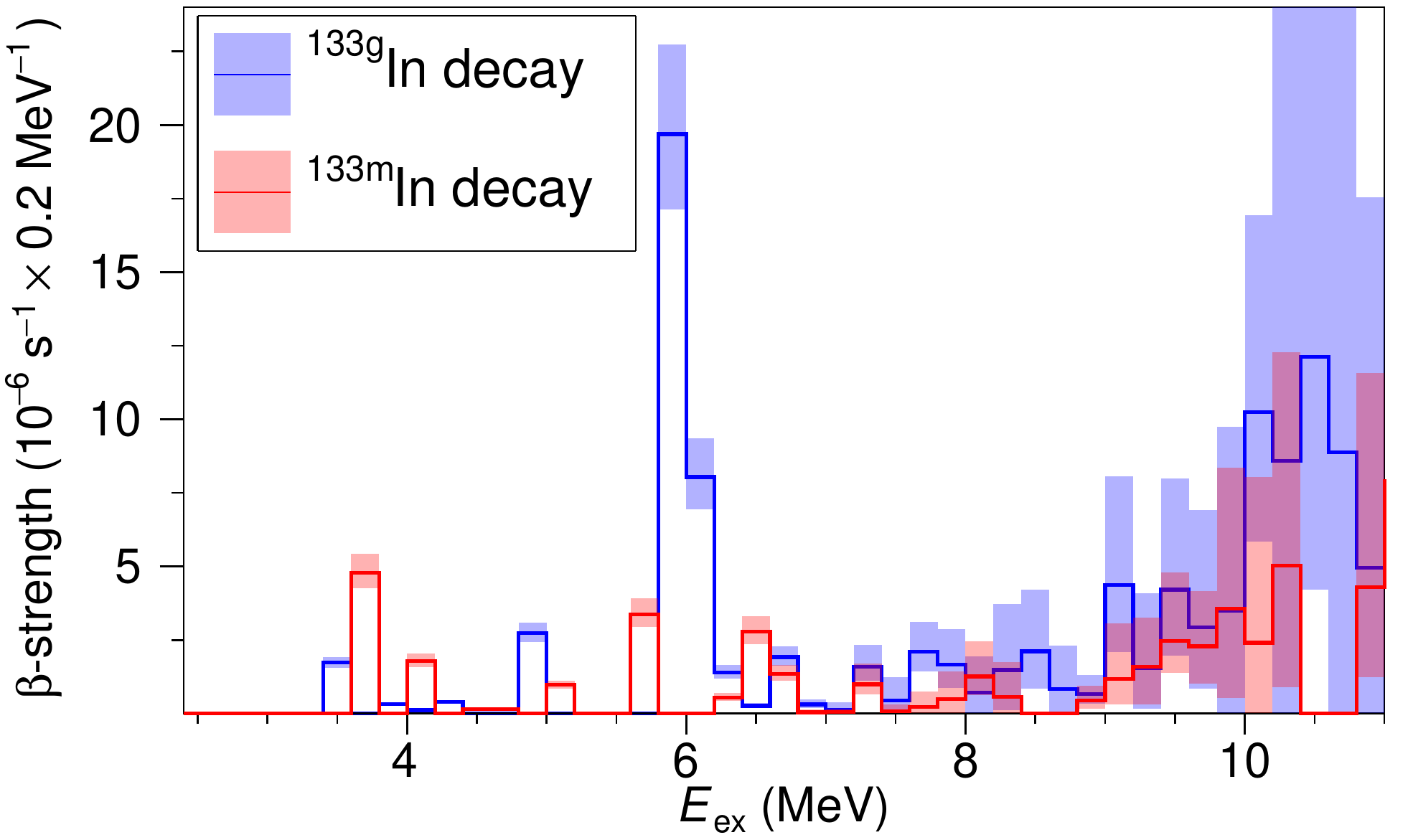}
   \caption{
      The experimental $\beta$-decay strength distribution (in $S_{\beta}=1/ft$
      \cite{sbeta}) of the $^{133}$In ground state (blue) and isomer (red)
      between $E_{ex}=2.5$ and 11 MeV.
   }\label{expsbeta}
\end{figure}

\Cref{expsbeta} presents the extracted $\beta$-strength distribution, in the
form of $S_{\beta}=1/ft$ \cite{sbeta}, of the $^{133g,m}$In decays with an
energy interval of 200 keV. The distributions include the states listed in
\cref{133Sn-table} and the decay strengths associated with the neutron emissions
feeding the $^{132}$Sn excited states, which is around 7\% of total neutron
emissions as discussed above. For these minor strengths, the neutron-$\gamma$
coincidence analysis was needed to correct the excitation energies of the
neutron-unbound states in $^{133}$Sn. This can be easily applied to the neutron
emissions that feed the $2^+$, $3^-$, or $4^+$ states in $^{132}$Sn due to their
relatively strong neutron-$\gamma$ cascades. Additionally, Piersa \textit{et
al}.\ and Benito \textit{et al}.\ observed weak neutron emissions feeding the
states higher than the $4^+$ state in $^{132}$Sn \cite{monika, benito}. Their
observations were confirmed in our measurement, but the associated $\gamma$
decays were generally too weak to perform credible neutron-$\gamma$ coincidence
analysis for the energy correction. To simplify the analysis and include their
contribution in \cref{expsbeta}, two extreme cases were considered. First, the
strength distribution was calculated assuming those weak neutron emissions only
fed the lowest $2^+$, $3^-$, or $4^+$ states around 4 MeV in $^{132}$Sn. Second,
the calculation was repeated with the $^{132}$Sn state shifted to 6.5 MeV, the
highest observed state in the $\beta$-delayed neutron emissions of $^{133g,m}$In
\cite{benito}. The final results shown in \cref{expsbeta} are the average
between the two calculations, with the error bars covering their upper and lower
limits.

\section{Spin and parity assignments}\label{spin}

Before this work, the only state known to have a neutron-hole configuration in
$^{133}$Sn was the $11/2^-$ state at 3564 keV \cite{hoff, monika, vaquero17}.
Its wavefunction is dominated by a neutron 2p-1h configuration, in which a
neutron hole at $h_{11/2}$ couples to two neutron particles above $N=82$. In the
$^{133g}$In decay, the same state was observed at $E_{ex}=3561(18)$ keV, in good
agreement with the literature value. It is the lowest neutron unbound state seen
in the experiment (see \cref{decayschemeP}). In an intuitive picture, the decay
is associated with a FF transition $\nu h_{11/2}\rightarrow\pi g_{9/2}$, during
which the two neutron particles outside $N=82$ persist as a spin $J=0$ pair in
the initial and final states. Fogelberg \textit{et al}.\ observed the analogous
transition without the neutron pair in $^{131}$In$\rightarrow^{131}$Sn with a
$\logft>5.6$ \cite{Sn-131}. According to the odd-mass tin isotopes with $N<82$,
there are two extra neutron orbitals $d_{3/2}$ and $s_{1/2}$ close to
$h_{11/2}$, giving rise to three neutron 2p-1h states at similar excitation
energy in $^{133}$Sn. Indeed, two states were observed in the isomeric decay at
excitation energies of 3.62 and 3.79 MeV. Following the systematics, the lower
state ($\logft=5.4$) was assigned $I^{\pi}=3/2^+$, and the upper state
($\logft=5.8$) $I^{\pi}=1/2^+$. Their underlying transitions are $\nu
d_{3/2}\rightarrow\pi p_{1/2}$ and $\nu s_{1/2}\rightarrow\pi p_{1/2}$,
respectively, both of which are FF transitions. The most important neutron-hole
orbital involved in the $^{133}$In decay is the deeply bound $\nu g_{7/2}$
because it determines the GT strength of $^{133g}$In. In \cref{expsbeta}, one
can see a remarkable strength at $E_{ex}=5.92$ MeV exclusive to the $^{133g}$In
decay. Its $I_{\beta}$ gives a $\logft=4.7$, which is significantly stronger
than any other feeding in the decay. Thus, the state was assigned
$I^{\pi}=7/2^+$ originating from $\nu g_{7/2}^{-1}$. The newly determined GT
strength of $\nu g_{7/2}\rightarrow\pi g_{9/2}$ in $^{133g}$In has a similar
$\logft$ as in the $^{131g}$In decay (=4.4 \cite{Sn-131}). It is noted that this
assignment is different from that in Ref.\ \cite{monika} suggesting the 6088-keV
state to be populated via the dominant GT decay.

\begin{figure}[htbp]
   \centering
   \includegraphics[width=0.5\textwidth]{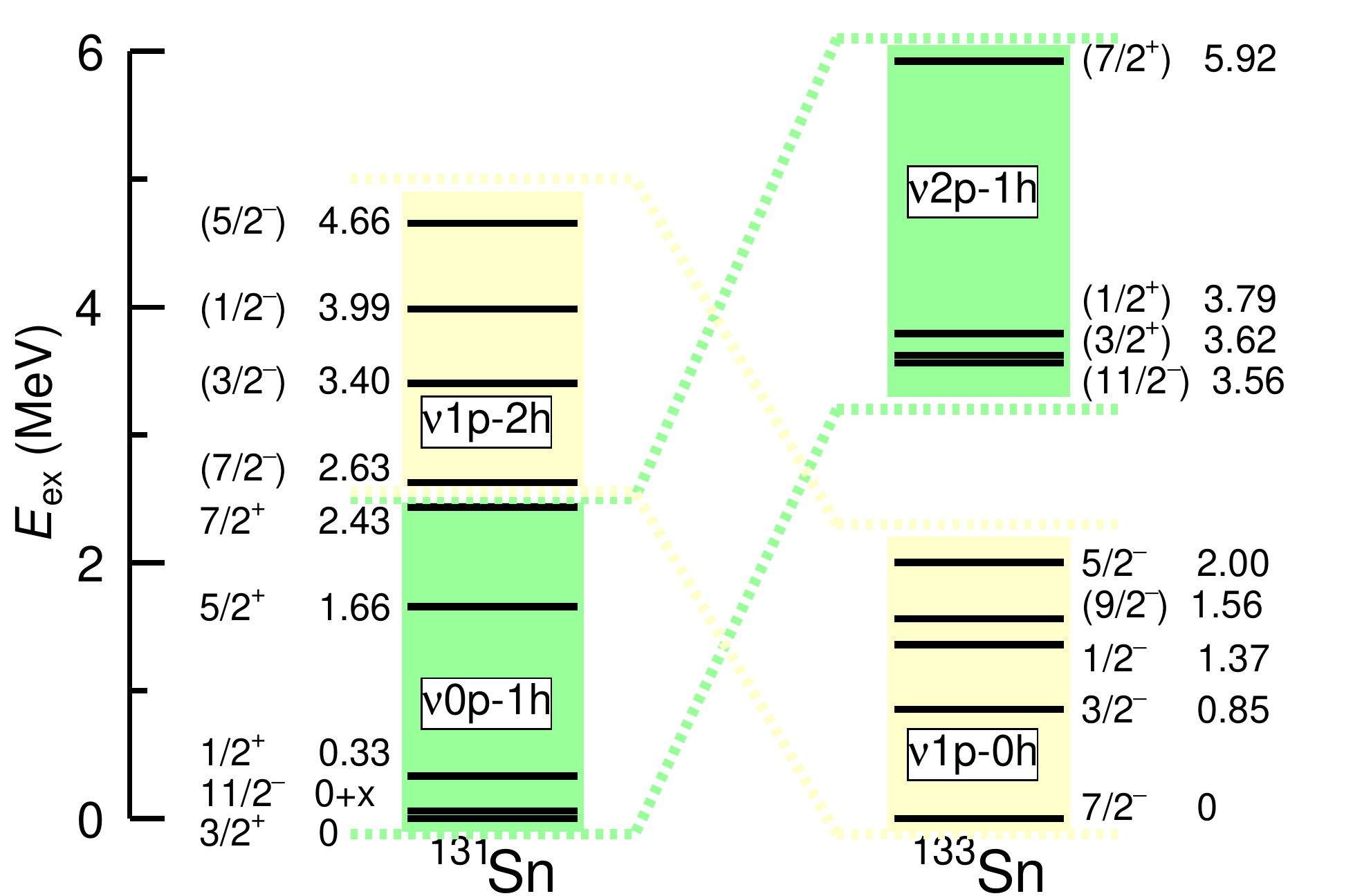}
   \caption{Reduced level schemes of $^{131}$Sn and $^{133}$Sn. Each group of
   states is labeled by the dominant neutron $\ph$ configuration in the wave
   functions. Information in $^{131}$Sn is taken from Refs.\ \cite{Sn-131,
   sn130dp}. In $^{133}$Sn, the spins, parities, and excitation energies of the
   $\nu$1p-0h states are taken from Refs.\ \cite{monika, allmond}, while those
   of $\nu$2p-1h are from this work. The dashed lines connect groups of states
   with the same odd-even combination of neutron p-h configuration.
   }\label{131-133-compare}
\end{figure}

The result enables a complete comparison between $^{131}$Sn and $^{133}$Sn, see
\cref{131-133-compare}. Before this work, the neutron 1p-2h states in $^{131}$Sn
and the neutron 1p-0h states in $^{133}$Sn had revealed remarkable similarity in
neutron transfer reactions \cite{133Sn_kate, sn130dp}. Similarly, if one moved
down the neutron 2p-1h states in $^{133}$Sn and aligned the lowest $11/2^-$
state to the ground state in $^{131}$Sn, the level scheme constructed in this
work is analogous to that of the 0p-1h states in $^{131}$Sn \cite{Sn-131},
supporting our spin-parity assignment discussed previously.

Besides, one notices strong $\beta$ feedings at $E_{ex}\sim5$ MeV in
\cref{expsbeta} in both ground-state and isomeric decays. At this excitation
energy, it is possible to break the proton $Z=50$ core and populate the proton
1p-1h excited states in $^{133}$Sn. From the low-lying states in odd-mass
antimony isotopes ($Z=51$) around $N=82$, $\pi g_{7/2}$ and $\pi d_{5/2}$ are
expected to be the lowest two proton orbitals outside $Z=50$. Thus, the lowest
proton core excited states observed in the $^{133}$In decay should be dominated
by either $\pi (g^{-1}_{9/2}\ g_{7/2}) \times\nu f_{7/2}$ or $\pi (p^{-1}_{1/2}\
g_{7/2})\times\nu f_{7/2}$, depending on where the proton hole is in the initial
state, i.e., whether the ground-state or isomeric decay. Both scenarios are
carried by the FF transition $\nu f_{7/2}\rightarrow \pi g_{7/2}$. At slightly
higher excitation energy, states with the $\pi(g^{-1}_{9/2}\ d_{5/2})\times\nu
f_{7/2}$ or $\pi(p^{-1}_{1/2}\ d_{5/2})\times\nu f_{7/2}$ configuration should
also be accessible via $\nu f_{7/2}\rightarrow\pi d_{5/2}$. Benito \textit{et
al}.\ observed analogous transitions in the decay from $^{132}$In to $^{132}$Sn
at similar excitation energy \cite{benito}. The $I^{\pi}$ of those states were
assigned based on the $\beta$-decay selection rules of non-unique FF
transitions: $\Delta I=0$ or 1 and $\Delta\pi=-1$. The unique FF transitions
with $\Delta I=2$ were not considered in the present spin assignment due to
their significantly larger $\logft$ and smaller intensities \cite{logftreview}.
For example, the neutron $d^{-1}_{5/2}$ state in $^{131}$Sn is populated in the
isomeric decay of $^{131}$In ($\nu d_{5/2}\rightarrow\pi p_{1/2}$) with a
$\logft=9.5$ \cite{Sn-131}, of which the feeding probability is far below our
sensitivity.

Above the $7/2^+$ state at 5.92 MeV, no isolated resonances were seen with
strong feeding strength in either of the $^{133g,m}$In decay. The spectra follow
a continuous distribution, which was attributed to the high level density and
limited resolving power. In this energy region, $\beta$ decay can populate both
positive- and negative-parity states via GT or FF transitions. The GT
transitions are favored because of their more significant matrix elements.
Following the selection rule, the states with $E_{ex}>6$ MeV were assigned the
spin $I=(7/2,~9/2,~11/2)$ in the ground-state decay and $(1/2,~3/2)$ in the
isomeric decay.

\section{Summary and conclusions}

In this work, the $\beta$-decay properties of $^{133g,m}$In were studied at IDS.
With the use of $\beta$-delayed $\gamma$ and neutron spectroscopy, their major
components in the decay-strength distribution were located above the neutron
separation energy in the daughter $^{133}$Sn. The strong GT transformation $\nu
g_{7/2}\rightarrow\pi g_{9/2}$ was observed in the $^{133g}$In decay, feeding a
$7/2^+$ state at 5.92 MeV in $^{133}$Sn. Besides, many neutron-unbound states
originating from neutron or proton $\ph$ excitations were found at lower
energies following the FF decays of $^{133g,m}$In. The spins and parities of
those states were assigned tentatively based on the $\beta$-decay selection
rules, the extracted $\logft$ values, and systematics along the isotopic chain.

The experimental findings greatly extend our knowledge of the $^{133}$In decay
from previous works \cite{hoff, monika, benito}, providing the $\beta$-strength
distribution southeast of $^{132}$Sn. The results are crucial to benchmark
$\beta$-decay theories and will serve as a bridge to understand the decay
properties of more neutron-rich nuclei, e.g., those $r$-process waiting-point
nuclei near the $N=82$ shell closure.

\begin{acknowledgments}
We acknowledge the support of the ISOLDE Collaboration and technical teams.
This project was supported by the European Unions Horizon 2020 research and
innovation programme Grant Agreements No.\ 654002 (ENSAR2),
by the Office of Nuclear Physics, U.S. Department of Energy under Award No.\
DE-FG02-96ER40983 (UTK) and DE-AC05-00OR22725 (ORNL),
by the National Nuclear Security Administration under the Stewardship Science
Academic Alliances program through DOE Award No.\ DE-NA0002132,
by the Romanian IFA project CERN-RO/ISOLDE,
by the Research Foundation Flanders (FWO, Belgium),
by the Interuniversity Attraction Poles Programme initiated by the Belgian
Science Policy Office (BriX network P7/12),
by the German BMBF under contracts 05P18PKCIA and 05P21PKCI1 in Verbundprojekte
05P2018 and 05P2021,
by the UK Science and Technology Facilities Research Council (STFC) of the UK
Grant No.\ ST/R004056/1, ST/P004598/1, ST/P003885/1, ST/V001027/1, and
ST/V001035/1,
by National Natural Science Foundation of China under Grant No.\ 11775316,
by the Polish National Science Center under Grants No.\ 2019/33/N/ST2/03023,
No.\ 2020/36/T/ST2/00547, and No.\ 2020/39/B/ST2/02346,
by Spanish MCIN/AEI FPA2015-65035-P, PGC2018-093636-B-I00, RTI2018-098868-B-I00,
PID2019-104390GB-I00, PID2019-104714GB-C21, and IJCI-2014-19172 grants,
by Universidad Complutense de Madrid (Spain) through Grupo de F\'isica Nuclear
(910059) and Predoctoral Grant No.\ CT27/16-CT28/16.
\end{acknowledgments}

%\bibliography{ref1,ref2}
%
\end{document}